\documentclass[preprint,number,sort&compress,3p,twocolumn]{elsarticle}
\usepackage{amsmath}
\usepackage[mathlines]{lineno}
\usepackage{times,helvet,courier}
\usepackage{a4g,graphicx}
\usepackage[bf]{caption}
\usepackage[bf]{subfigure}
\usepackage{wasysym}
\usepackage{float}
\usepackage[T1]{fontenc}
\usepackage{epstopdf}
\begin{document}

\hyphenation{COBRA}

\begin{frontmatter}

\title{Analytical model for event reconstruction in coplanar grid CdZnTe detectors}

\author[Dresden]{Matthew Fritts\corref{cor1}}
\ead{matthew\_christopher.fritts@tu-dresden.de}
\author[Erlangen]{J\"urgen Durst}
\ead{juergen.durst@physik.uni-erlangen.de}
\author[Dresden]{Thomas G\"opfert}
\ead{thomas.goepfert@tu-dresden.de}
\author[Dresden]{Thomas Wester}
\ead{Thomas.Wester@physik.tu-dresden.de}
\author[Dresden]{Kai Zuber}
\ead{zuber@physik.tu-dresden.de}
\address[Dresden]{Institut f\"ur Kern- und Teilchen-Physik (IKTP), Technische Universit\"at Dresden, D-01062 Dresden, Germany}
\address[Erlangen]{Erlangen Centre for Astroparticle Physics (ECAP), 
Friedrich-Alexander Universit\"at Erlangen-N\"urnberg, 
Erwin-Rommel-Str. 1, D-91058 Erlangen, Germany}
\cortext[cor1]{Corresponding author. Tel 49 351 46334568; fax 49 351 46337292.}

\begin{abstract}
Coplanar-grid (CPG) particle detectors were designed for materials such as CdZnTe (CZT) in which 
charge carriers of only one sign have acceptable transport properties. The presence of two independent 
anode signals allows for a reconstruction of deposited energy based on the difference between the two 
signals, and a reconstruction of the interaction depth based on the ratio of the amplitudes of the sum 
and difference of the signals. Energy resolution is greatly improved by modifying the difference signal 
with an empirically determined weighting factor to correct for the effects of electron trapping. In this 
paper is introduced a modified interaction depth reconstruction formula which corrects for electron 
trapping utilizing the same weighting factor used for energy reconstruction. The improvement of this 
depth reconstruction over simpler formulas is demonstrated. Further corrections due to the contribution 
of hole transport to the signals are discussed.
\end{abstract}


\end{frontmatter}


\section{Introduction}

CZT as a $\gamma$-ray detector material has a number of attractive properties such as a large band-gap and good 
performance at room temperature. The COBRA experiment~\cite{cobra}
uses CZT detectors in a search for neutrinoless 
double-beta decay ($0\nu\beta\beta$) due to the presence of several $0\nu\beta\beta$ candidate isotopes in 
CdZnTe and its low natural background radioactivity. 
Currently COBRA utilizes a $4\times4\times2$ array of 1~cm$^{3}$ CPG detectors 
operating under low background conditions at the Gran Sasso National Laboratory (LNGS), with plans to upgrade 
to 64 detectors in early 2013.

Good energy resolution is an important consideration in gamma spectroscopy. It is also key for COBRA in order to enhance 
the sensitivity to a $0\nu\beta\beta$ line (produced by the summed energy of the two betas emitted) 
and to distinguish it from other backgrounds, including the continuous spectrum 
of the competing process of neutrino-accompanied double-beta decay ($2\nu\beta\beta$). The depth information provided by the 
CPG design is also very important, for two reasons. First, an important category of background is $\alpha$- or 
$\beta$-radiation at the cathode and anode surfaces. The short penetration of such background allows for its efficient 
discrimination, if the interaction depth is well known. Second, for certain regions in the detector distortion of 
the energy reconstruction is an unavoidable result of the CPG design. Depth information can be used to identify 
events in these regions and remove them from consideration as $0\nu\beta\beta$ candidates. For both of these phenomena analysis 
cuts are necessary, and an accurate assessment of the selection efficiency of such cuts requires the most accurate 
depth information achievable.

Such considerations give strong motivation for an accurate depth reconstruction for CPG detectors for their 
use in the COBRA experiment, and for all applications in which accurate knowledge of the active volume is 
required. In addition, due to the large number of detectors used in COBRA, an automatic method for generating 
depth formulas unique to each detector is highly desirable. Material properties and optimal operational parameters differ 
significantly from detector to detector, necessitating unique formulas for both energy and depth reconstruction.

\section{CPG principles}

The CPG detector design was introduced by P.~Luke in 1994~\cite{cpg94}. A proper understanding of the design principles begins 
with a treatment based on the Shockley-Ramo theorem. For a more detailed description of the application of the Shockley-Ramo
theorem to CPG detectors see~\cite{HeRamo}.

\begin{figure}
\centering
\includegraphics[width=0.5\textwidth]{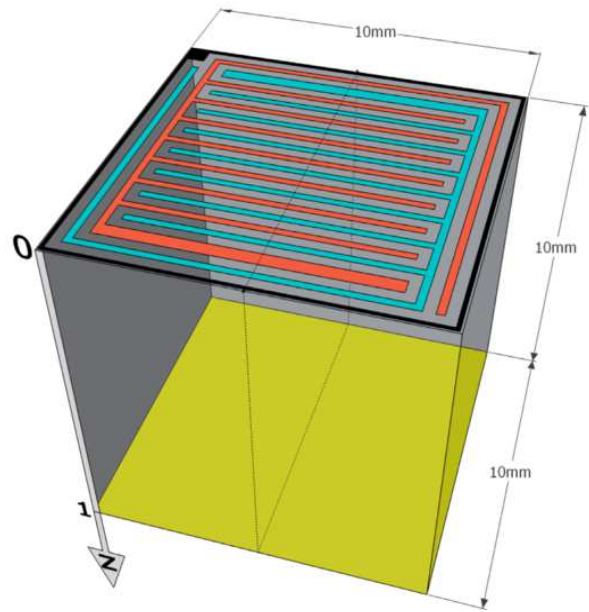}
\caption{Schematic electrode layout for CPG detectors used by COBRA. Two interleaved anodes and an outer 
guard ring (uninstrumented) are patterned on one side (top); the other side (bottom) is a uniform cathode. Shown
here is the convention used in this paper for the $z$ axis (interaction depth). The dashed lines indicate the
section over which the calculated weighting potential is shown in Figure~\ref{fig-Vw2d}. }
\label{fig-detector}
\end{figure}

In a CPG detector such as those used in COBRA, the cathode is patterned on one side as a uniform rectangle, while on 
the opposite side two isolated anode grids are formed in the shape of interlocking combs. Figure~\ref{fig-detector} 
is a schematic of a 
COBRA CPG detector, manufactured by EI Detection \& Imaging Systems~\cite{endicott}. 
In operation a large (approximately 1~kV) bias (referred to as HV) is 
applied between the cathode and one anode, while a smaller (typically 50-100~V) bias (GB) is applied between the anode grids. 
The anode held at higher potential is referred to as the collecting anode (CA); the other anode is known as the non-collecting 
anode (NCA). Electrons excited into the conduction band 
by a particle interaction somewhere in the bulk of the detector drift straight toward the 
anode plane (in the negative-$z$ direction) until they get close to the anode grid rails, at which time they are 
diverted by the near-anode field to be collected in the CA.

\begin{figure}
\centering
\includegraphics[width=0.5\textwidth]{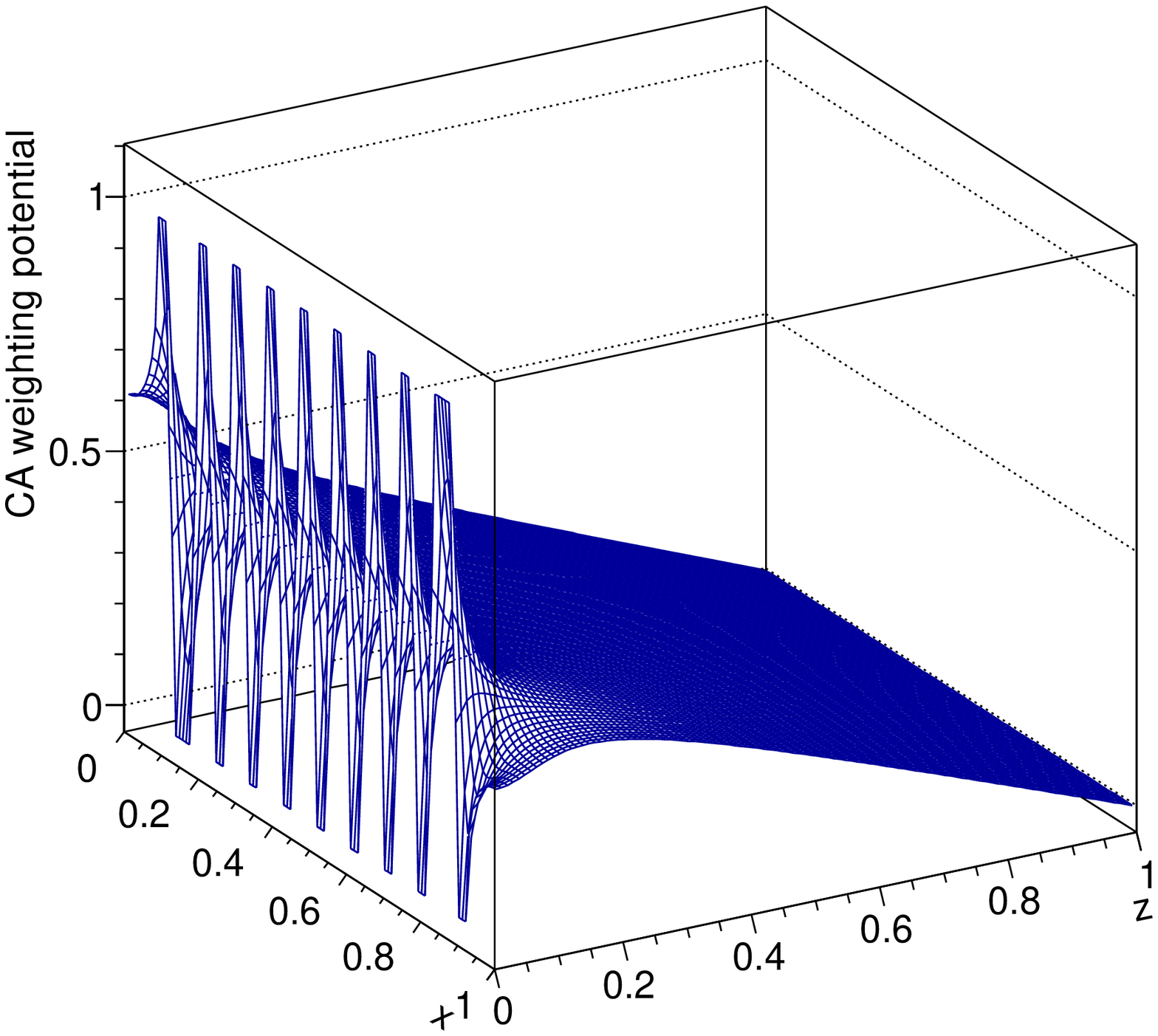}
\includegraphics[width=0.5\textwidth]{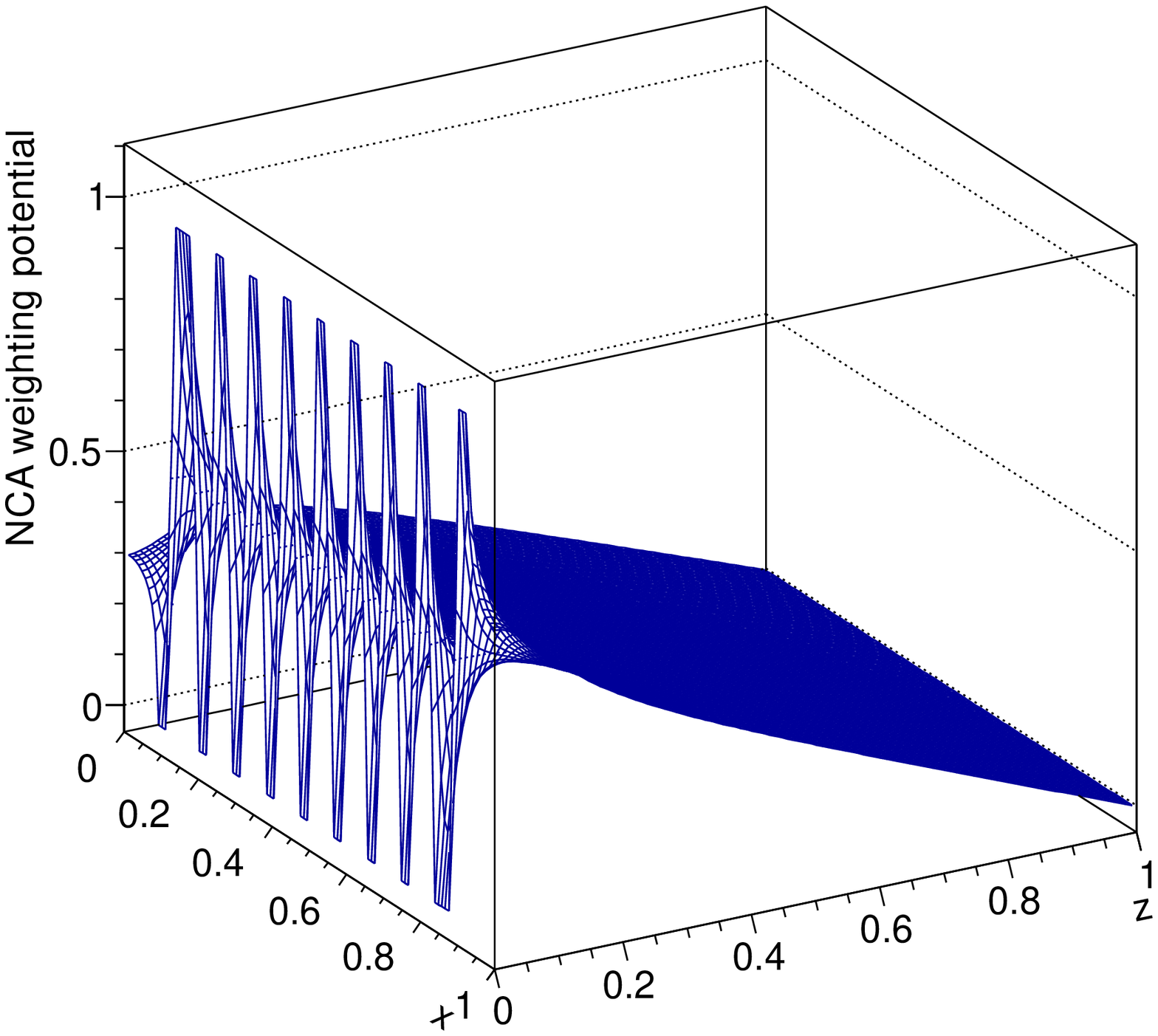}
\caption{Weighting potentials of CA and NCA as calculated for the COBRA CPG detector geometry 
across a plane through the center of the detector. The coordinates are normalized to the detector dimensions, 
thus ranging from 0 to 1.}
\label{fig-Vw2d}
\end{figure}

A first step in the application of the Shockley-Ramo theorem is to calculate the weighting potentials~\cite[p.~813-8]{hecht}. 
Figure~\ref{fig-Vw2d} shows the weighting potentials of the CA and NCA along a plane through the center of the detector,
 calculated using the electrode geometry of COBRA detectors.\footnote{
In current operation the guard ring electrode (black in Figure~\ref{fig-detector}) is left unconnected. This
information is incorporated into the weighting potential calculations.}
At most locations in the detector, the two
weighting potentials are nearly equal, rising with a slope of 1/2 from the cathode plane to the anode plane.
At the position of a CA rail (normally the final location of the mobile electrons created by ionization), 
the CA and NCA weighting
potentials are 1 and 0, respectively, by definition. 

These properties of the weighting potential allow for a
difference signal based on the raw CA and NCA signals that is proportional to the charge from ionization and 
independent of the interaction depth. From the Shockley-Ramo theorem one can calculate the amplitudes of
the two anode signals produced by electrons:
\begin{linenomath*}
\begin{equation}
\Delta q_{CA} = \frac{1}{2} Q_0 ( z_0 + 1 )
\label{eq-CA0}
\end{equation}
\begin{equation}
\Delta q_{NCA} = \frac{1}{2} Q_0 ( z_0 - 1 )
\label{eq-NCA0}
\end{equation}
\end{linenomath*}
Here $\Delta q$ represents the change in induced charge, and thus the amplitude of the signal, for the 
corresponding anode. $Q_0$ is the magnitude of the mobile charge produced by the interaction.  $z_0$ is the distance
 between the interaction location and the anode plane expressed as a dimensionless fraction of the detector
length. $z_0$ is referred 
to as the ``interaction depth'' (or simply ``depth''). 

The depth dependence in the two raw anode signals is removed by subtraction:
\begin{linenomath*}
\begin{equation}
\Delta q_{CA} - \Delta q_{NCA} = Q_0
\end{equation}
\end{linenomath*}
Note that $\Delta q_{NCA}$ is always negative, so the equation represents a sum of the absolute amplitudes. 
By calibrating this signal using sources of known spectra one achieves a well-resolved reconstruction 
of the energy deposited by an interaction.

Similarly one can remove the charge magnitude dependence to reconstruct the interaction depth~\cite{HeDepth}:
\begin{linenomath*}
\begin{equation}
\frac{\Delta q_{CA} + \Delta q_{NCA}}{\Delta q_{CA} - \Delta q_{NCA}} = z_0
\label{eq-z0}
\end{equation}
\end{linenomath*}

 The picture of CPG operational principles presented in this section can be referred to as 
zeroth-order behavior. First-order effects, which complicate this simple picture, are the subject of the following sections.

\section{Electron Trapping}

Electron trapping in CZT detectors is a strong enough effect that, for COBRA detectors, the signal amplitude of an 
interaction occurring at large depth is reduced by about 10\% relative to that of an interaction occurring at small 
depth. This depth dependence would significantly degrade energy resolution were it not corrected. Such a correction is 
possible due to the depth information intrinsic in the raw signals.

P.~Luke and E.~Eissler introduced a simple weighting factor (referred to here as $w$) to modify the difference signal in order 
to correct for electron trapping~\cite{weighting-factor}:
\begin{linenomath*}
\begin{equation}
\Delta q_{CA} - w\Delta q_{NCA}
\end{equation}
\end{linenomath*}
The weighting factor is empirically determined and is always less than one. The correction for trapping can be 
understood qualitatively: as interaction depth increases, $-\Delta q_{NCA}$ becomes smaller while the signal reduction 
due to trapping becomes larger. Underweighting the NCA amplitude thus artificially lowers the 
amplitudes of low-depth interactions to better match the trapping-degraded amplitudes of high-depth interactions.

\begin{figure}
\centering
\includegraphics[width=0.5\textwidth]{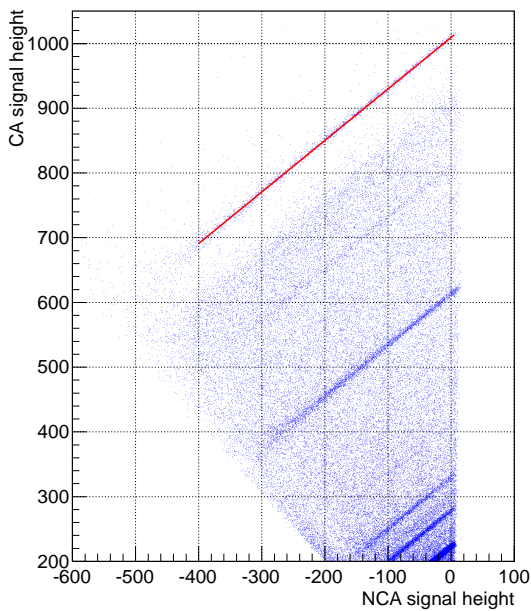}
\caption{CA versus NCA signal heights for $^{228}$Th calibration data for one COBRA detector. $\gamma$-lines 
from the source appear as upward-sloping lines in this plot. The red line indicates 2.615~MeV events from $^{208}$Tl decay. 
The slope of this line, 0.80, 
is the empirically-determined weighting factor $w$. In the absence of trapping this slope would be 1.}
\label{fig-CAvsNCA}
\end{figure}

The weighting factor $w$ can be determined by an optimization procedure seeking the best energy resolution. 
Alternatively if information from both signals is recorded separately the weighting factor can be determined 
from a single calibration run by analysis of the CA and NCA signal amplitudes for a calibration line. 
Figure~\ref{fig-CAvsNCA} illustrates this procedure as performed in COBRA.

In the following paragraphs an analytical treatment of the electron trapping effect in CPG detectors is presented,
using a few simplifying idealizations. This treatment is similar to the derivation in the Appendix of~\cite{luke2007}, 
although here the anode signals are separately derived. 
First assume a mean trapping length $\lambda$ that is valid for the 
entire electron drift path.
This is only true to the extent that both the crystal properties and the electric field 
are uniform throughout the detector volume. In actuality the electric field is nearly uniform for a large majority 
of the electron path for most interaction locations, and thus for most of the drift distance in which trapping 
occurs.\footnote{This is true considering the electrode geometry if one assumes that space charge has a negligible effect.
The shape of charge pulses in COBRA detectors show no evidence of significant field distortions due to space charge.}
Thus the magnitude of the charge as a function of drift distance $d$ is taken to be
\begin{linenomath*}
\begin{equation}
Q = Q_0 e^{-d/\lambda}
\end{equation}
\end{linenomath*}
As with $z_0$, $\lambda$ and $d$ are expressed in terms of detector length and are thus dimensionless.
$\lambda$ is considered a free parameter in this treatment. Physically it depends upon the electron mobility-lifetime
 product for the crystal and the voltage applied to the crystal, both of which vary by detector. 
Thus $\lambda$ is detector dependent.

\begin{figure}
\centering
\includegraphics[width=0.5\textwidth]{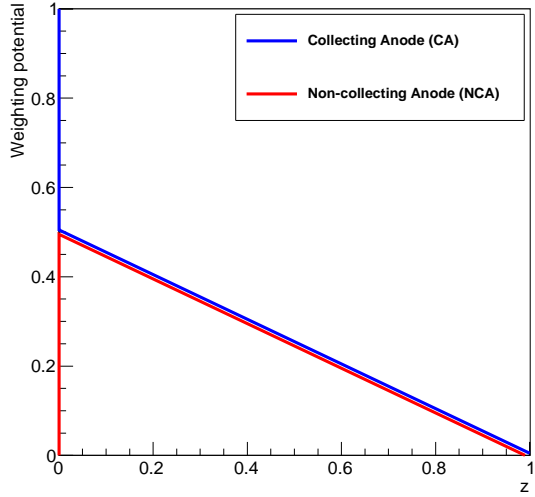}
\caption{Idealized weighting potentials of CA and NCA corresponding to path of charges drifting through a CPG detector.}
\label{fig-Vw1d}
\end{figure}

A second simplification is to assume that the weighting potentials of both anodes take their bulk form along the
entire electron drift path up until the anode plane is reached, at which point the CA and NCA weighting potentials 
abruptly change to 1 and 0, respectively. The idealized weighting potentials as functions of depth $z$ are illustrated 
in Figure~\ref{fig-Vw1d}. 
In this model the electron drift is reduced to one dimension, with the
 drift entirely in the negative $z$ direction. For most interaction
 locations this is a reasonable approximation in that, again, most of the trapping occurs before deviations in the weighting
 potential become significant and before the electron path is diverted by the near-grid field. The expected size of the error
 introduced by this approximation is discussed in a later section.

Given these assumptions the expected anode signal amplitudes can be calculated. We divide the calculation into two parts. 
The first part corresponds to the electrons drifting through the region of uniform weighting potentials, where
$\frac{dV_w}{dz}=-\frac{1}{2}$. Since the charge 
magnitude changes along the drift path we calculate this with an integral form of the Ramo equation, yielding
\begin{linenomath*}
\begin{equation}
\begin{split}
\Delta q_1& = \int_0^{z_0} \frac{1}{2} Q_0 e^{-\left(z_0-z\right)/\lambda} dz \\
&= \frac{1}{2} Q_0 \lambda \left( 1 - e^{-z_0/\lambda} \right) \\
\end{split}
\end{equation}
\end{linenomath*}
This result is valid for both the CA and NCA signals. It is closely related to the Hecht equation~\cite[p.~489]{hecht}, differing only by 
the factor one-half. In a signal formed by the sum of the CA and NCA signals (sometimes referred to as the cathode signal) 
the exact Hecht equation is recovered. This corresponds to the weighting potential of a simple parallel-plate electrode geometry.

The second part of the calculation of the anode signal amplitudes corresponds to the abrupt changes in 
weighting potential at z=0: +1/2 for CA and -1/2 for NCA. Inserting the final charge into the Ramo equation one finds
\begin{linenomath*}
\begin{equation}
\Delta q_{2,CA} = \frac{1}{2} Q_0 e^{-z_0/\lambda}
\end{equation}
\begin{equation}
\Delta q_{2,NCA} = -\frac{1}{2} Q_0 e^{-z_0/\lambda}
\end{equation}
\end{linenomath*}

The full expressions for the anode signals are
\begin{linenomath*}
\begin{equation}
\Delta q_{CA} = \frac{1}{2} Q_0 \left[ \lambda \left( 1 - e^{-z_0/\lambda} \right) + e^{-z_0/\lambda} \right]
\label{eq-CAtrapping}
\end{equation}
\begin{equation}
\Delta q_{NCA} = \frac{1}{2} Q_0 \left[ \lambda \left( 1 - e^{-z_0/\lambda} \right) - e^{-z_0/\lambda} \right]
\label{eq-NCAtrapping}
\end{equation}
\end{linenomath*}

By inspection we see that, in the limit of large $\lambda$, the corresponding no-trapping forms 
(Equations~\ref{eq-CA0} and \ref{eq-NCA0}) are restored. 

As in the zeroth-order case the position dependence can be removed by a linear combination of the two signals to reconstruct 
the charge excited into the conduction band by the interaction:
\begin{linenomath*}
\begin{equation}
\Delta q_{CA} - \frac{\lambda -1}{\lambda+1} \Delta q_{NCA} = \frac{\lambda}{\lambda +1} Q_0
\label{eq-wdiff-trapping}
\end{equation}
\end{linenomath*}
It is noteworthy that the trapping effects, themselves of exponential form, are canceled exactly with a linear combination.
For this it is necessary only that the $z_0$-dependent terms in equations \ref{eq-CAtrapping} and \ref{eq-NCAtrapping} are 
identical in form.
On the left-hand side of equation~\ref{eq-wdiff-trapping} is a weighted difference signal equivalent to the one introduced by Luke. On the right-hand side is 
the excited charge reduced by a constant factor, as expected from the qualitative understanding of the weighted difference. 
The constant is of course ultimately absorbed into the energy calibration. The relationship between the parameter $\lambda$ 
and the empirical weighting factor $w$ is
\begin{linenomath*}
\begin{equation}
\lambda = \frac{1+w}{1-w}
\end{equation}
\end{linenomath*}
An equivalent relationship is derived in~\cite{luke2007}.

It is also possible to construct an analytical expression for the interaction depth $z_0$ from the raw signal formulas
by eliminating $Q_0$ from equations \ref{eq-CAtrapping} and \ref{eq-NCAtrapping}:
\begin{linenomath*}
\begin{equation}
\lambda \ln \left( 1 + \frac{1}{\lambda} \frac{\Delta q_{CA} + \Delta q_{NCA}}{\Delta q_{CA} - \Delta q_{NCA}}  \right) = z_0
\end{equation}
\end{linenomath*}

Note that as expected the zeroth-order depth formula is recovered in the limit of large $\lambda$. This trapping-corrected 
depth formula is easily implemented because the parameter $\lambda$ can be calculated from $w$ which is already determined 
during the energy correction procedure.

\section{Evaluation of the trapping model}

In the framework of the simple model for trapping presented here we can predict the performance of the 
zeroth-order depth formula, now seen as an approximation. It is clear that it is non-linear; moreover, the 
zeroth-order formula will overestimate the interaction depth in a detector-dependent way. From the empirical 
weighting factors $w$ determined for COBRA detectors the overestimation can be expected to reach a maximum of 
around 3-10\% for $z_0=1$.

\begin{figure}
\centering
\includegraphics[width=0.5\textwidth]{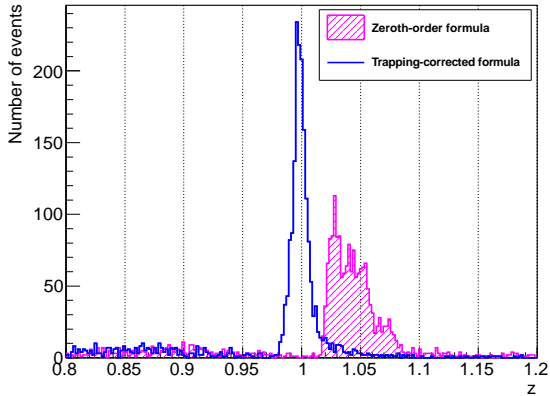}
\caption{Interaction depth $z$ of near-cathode events in the COBRA data collected at LNGS in 31 detectors 
with deposited energies greater than 2 MeV. 
An excess of interactions occurring very near the cathode is known to exist. Uncorrected depth reconstruction results in depths overestimated
by a few percent and a wider distribution due to detector differences.}
\label{fig-ztcVSz0}
\end{figure}

Figure~\ref{fig-ztcVSz0} illustrates the advantages of the new depth formula. 
Sources of alpha radiation at the cathode 
surfaces for the detectors currently in operation at LNGS produce an excess of events near $z=1$. The true 
interaction depth for these events is expected to be very close to 1. The figure compares the depth distribution 
using both the zeroth-order and trapping-corrected formulas. The predicted advantages are evident: the corrected 
depths are much closer to one, and they are much more tightly distributed.

\begin{figure}
\centering
\includegraphics[width=0.5\textwidth]{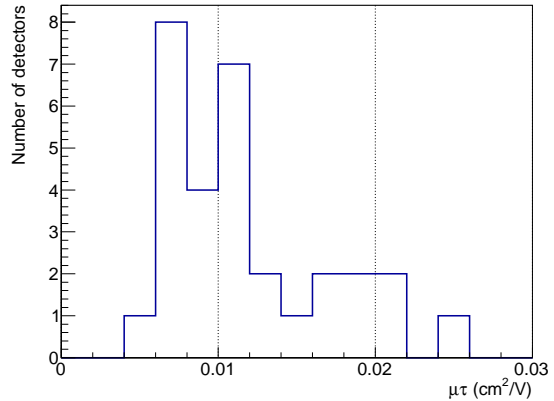}
\caption{Electron mobility-lifetime products of 30 COBRA detectors, estimated by the method described in the text. The mean
value is $1.1 \times 10^{-2}$  cm$^{2}$ / V.  }
\label{fig-mutau}
\end{figure}

The empirically determined parameter $\lambda$ can be used to estimate the electron mobility-lifetime 
product ($\mu \tau$) for each detector as follows:
\begin{linenomath*}
\begin{equation}
\mu \tau = \frac{L^2  \lambda}{V_{bulk}}
\end{equation}
\end{linenomath*}
where $L$ is the detector length (1~cm) and $V_{bulk}$ is calculated to be the potential difference 
between the cathode and the average of the anode potentials (HV - $\frac{1}{2}$GB). Figure~\ref{fig-mutau}
 shows the resulting $\mu \tau$ estimates for 30 
COBRA detectors.  
The mean value of  
$1.1 \times 10^{-2}$~cm$^{2}$~/~V is close to those quoted for CZT in several publications (see Table 7 in \cite{mutau}).
However the wide distribution of values 
suggests significant variation across detectors, and possibly
 additional effects unaccounted for by this simplified model.\footnote{
One potential effect is an imbalance in the CA and NCA signal amplification factors arising from the specific 
properties of the electronics. If such an 
imbalance exists it will contribute to the weighting factor $w$ independently of electron transport properties.
However for this study such imbalances were measured and the $\lambda$ and $\mu \tau$ values corrected
accordingly. Neither can temperature be a factor since all detectors are operated under essentially identical
conditions at room temperature.}

\section{Additional first-order effects}

\subsection{Hole drift}

In CZT holes have much poorer transport properties than electrons; nevertheless they contribute 
significantly to the raw signal amplitudes. Because holes drift much more slowly than electrons, 
it is first necessary to consider whether their contributions to signal amplitudes are fully 
recorded in the data processing procedure. In standard COBRA pulse processing the effective time 
window within which the amplitude is measured is greater than 3~$\mu$s. The lifetime of holes 
in CZT is estimated at 1~$\mu$s, so that more than 95\% of the hole charge is trapped within the 
processing time window---assuming that the holes do not reach an electrode within this time.
Thus to a reasonable approximation we can assume holes contribute fully to the processed 
signal amplitudes.

In analogy to the parameter $\lambda$ in the treatment of electron trapping, we take $\rho$ to 
be the mean trapping length for holes, normalized to the detector length. Because typical $\lambda$ 
values are around 10, and because the mobility-lifetime product for holes in CZT is estimated to be 
about 1\% that of electrons, $\rho$ is estimated to be around 0.1. Thus for many interaction depths 
we do not expect holes to drift far enough to be collected at an electrode; rather they will be almost 
completely trapped after drifting a fraction of the detector length.

Qualitatively the hole effect can be pictured as follows: holes will drift a short distance toward the
 cathode, along a path in the bulk region of the weighting potential. They thus contribute equal positive
 contributions to the CA and NCA signal amplitudes. Since the energy reconstruction is based on the difference 
of these signals, the hole contributions will cancel each other. On the other hand the depth reconstruction is based on the sum of the 
anode signals, and so the hole effect can lead to a significant overestimation of interaction depth.

Quantitatively we can consider the hole effect using the same framework as was used for electron trapping. 
Assuming that the interaction depth occurs in the bulk region of the weighting potential, we find
\begin{linenomath*}
 \begin{equation}
\Delta q_{holes} = \frac{1}{2} Q_0 \rho \left( 1 - e^{-(1-z_0)/\rho} \right)
\end{equation}
\end{linenomath*}
This contribution applies to both anode signals. Considering the small value of  $\rho$ we see by inspection 
that the hole contribution is approximately $\frac{1}{2} Q_0 \rho$ for interaction depths far from the cathode 
(where $1-z_0$ is several times $\rho$). In contrast the expression falls to zero as $z_0$ approaches 1, since for near-cathode interactions the
 holes have no distance to drift. Therefore the hole contribution is zero for $z_0=1$. 

\begin{figure}
\centering
\includegraphics[width=0.5\textwidth]{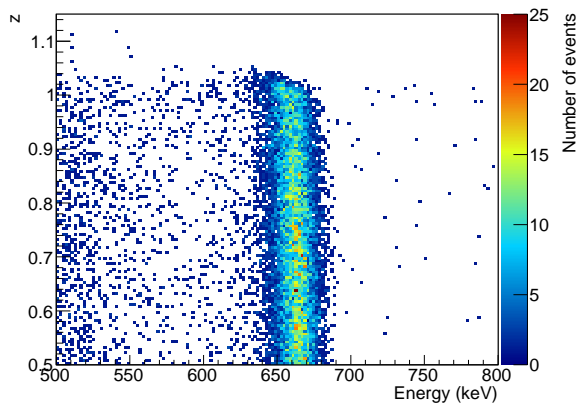}
\caption{Distribution of events in depth and energy in a COBRA detector irradiated with a $^{137}$Cs source. The reconstructed energy 
of 662 keV $\gamma$ events is somewhat underestimated at depths near 1 due to the hole effect leading to a deformation of the line 
near the top of the plot.}
\label{fig-Edistort}
\end{figure}

The contribution of holes to the signals has a second-order effect on the energy reconstruction,
producing a small reduction for near-cathode events~\cite{weighting-factor,luke2007}. For interactions 
far from the cathode the weighted difference signal (Equation~\ref{eq-wdiff-trapping}) 
is enhanced by approximately $\rho/\lambda$, on the 
order of 1\%. This constant factor will be absorbed into the energy calibration. However the hole enhancement 
disappears for interactions very close to the cathode. Thus the energy depositon of near-cathode events will be 
underestimated by about 1\%. This is comparable to the energy resolution of COBRA detectors. 
The effect is noticeable only in a layer whose thickness is a few percent of the detector length. Figure~\ref{fig-Edistort} 
shows this small effect on a calibration line with roughly uniform depth distribution.

\begin{figure}
\centering
\includegraphics[width=0.5\textwidth]{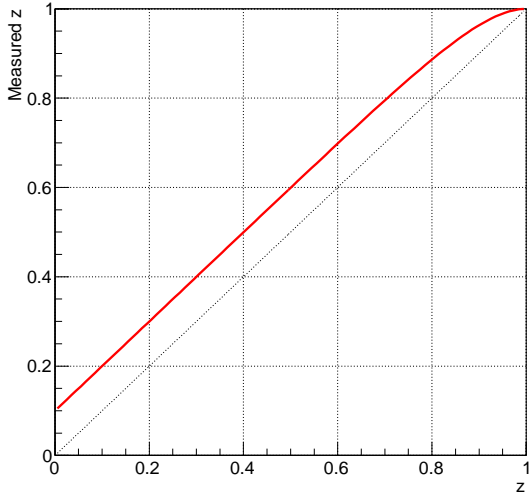}
\caption{Relationship between measured depth and true depth (red) based on an analytical 
treatment of the hole effect, assuming $\rho=0.1$. The measured depth is 
corrected for electron trapping but not corrected for the hole effect. }
\label{fig-ztcVSztrue}
\end{figure}

\begin{figure}
\centering
\includegraphics[width=0.5\textwidth]{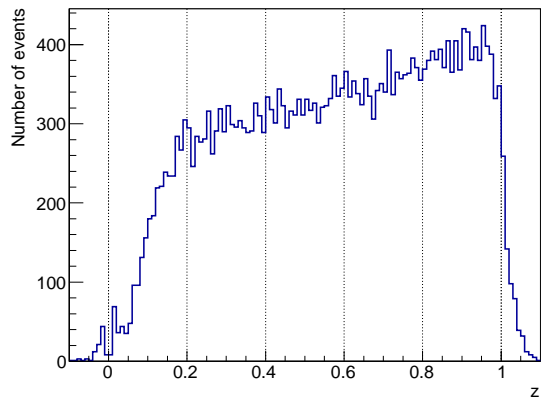}
\caption{Depth distribution of events from a COBRA detector for data taken at LNGS in an energy range of 180-300~keV. The events consist
almost entirely of intrinsic $\beta$-decay of $^{113}$Cd, and thus are known to be evenly distributed throughout the detector (although
some events near the detector surfaces will be lost due to escaping $\beta$ particles). The low count of measured depths below about 0.1 is
evidence of the hole effect. The hole effect is also partially responsible for the nonuniform distribution at greater depths. Distortions in energy-
and depth-reconstruction at lower depths due to the details of the weighting potential also contribute to this nonuniformity. }
\label{fig-z-dropoff}
\end{figure}

The hole effect on depth reconstruction is larger. It can be adequately described by considering the addition 
of the hole contribution to the zeroth-order equations (Equation~\ref{eq-CA0} and \ref{eq-NCA0}) and the corresponding 
effect on the zeroth-order 
depth formula (Equation~\ref{eq-z0}). For interactions far from the cathode, the depth will be overestimated by approximately $\rho$. 
For near-cathode interactions there is no overestimation, as is clear from Figure~\ref{fig-ztcVSz0}. 
Figure~\ref{fig-ztcVSztrue} illustrates the relationship 
between true and measured depth. The clearest evidence of this effect from data is the small number of events below a certain measured
 depth for populations of interactions known to have uniform depth distributions, as exemplified in Figure~\ref{fig-z-dropoff}. 

In principle of course one can also correct for the hole effect in the depth reconstruction. However it is more difficult to 
calculate the parameter $\rho$ in the calibration process. Another difficulty is that the analytic expression for 
hole-corrected depth has no solution if the zeroth-order depth is greater than 1. Since the finite resolution of 
measured depth sometimes results in values greater than 1, this is problematic for an event-by-event hole correction
algorithm.

\subsection{Near-anode distortions}

CPG event reconstruction depends on the nearly equal CA and NCA weighting potentials at the interaction point. However the 
two weighting potentials differ significantly at locations near the anode plane, as illustrated in Figure ~\ref{fig-Vw2d}.

The corresponding effects are somewhat complicated, and depend not only on depth $z_0$ but also on the $x$ and $y$ coordinates 
of the interaction point. The energy can be either underestimated or overestimated depending on whether the interaction point lies above 
a NCA or a CA rail. The thickness of the layer near the anode in which these effects are significant is greater near the lateral surface of 
the detector\footnote
{The distortions in the weighting potential can be reduced if a bias is applied to the guard ring~\cite{grid-improvement}. However for 
simplicity COBRA detectors are currently operated with the guard ring unconnected.} 
than it is near the center of the anode plane, where it is smaller than 0.5~mm. Where it is thinnest, holes can actually cancel 
the distorting effect, by drifting toward the cathode into the bulk region of the weighting potential.

Where the effective near-anode layer is thickest, holes will not drift far enough to cancel the distorting effects. As with the reconstructed
energy, the reconstructed depth is also
 distorted, but it will generally be smaller than $\rho$, so that a judicious depth selection can remove such events from consideration in data analysis.

Very near the anode the electric field in the detector is dominated by the grid bias, which produces a different effect. The region 
of effect depends strongly on the applied bias, but is estimated to be typically around 0.1~mm. For interactions in this region the
 holes will drift toward the NCA instead of the cathode. The corresponding hole contributions to the signals will be opposite in sign, 
thus strongly enhancing the difference signal. Due to the short hole drift distance, most holes are collected before they are trapped. 
The shape of the weighting potentials corresponding to the paths of both electrons and holes differ significantly from those corresponding
to events in the bulk. The net effect is that the reconstructed energy will be approximately doubled~\cite{HeDepth}. The reconstructed depth
will be very close to zero.

Signals from interactions far from the anode are also affected by the detailed form of the weighting potentials near the anode. 
If such effects are considered Equations \ref{eq-CAtrapping} and \ref{eq-NCAtrapping} 
must be corrected. This correction will vary with the $x$ and $y$ 
positions of the interaction. However we can approximate the size of the correction by considering a specific case: an interaction 
occurring on a line leading to the center of a CA rail in the central region of the anode plane. In this case the weighting 
potential as a function of $z$ fits well to an exponential form with characteristic length $\alpha$ of approximately 
0.015 (0.15 mm). We find then that Equations \ref{eq-CAtrapping} and \ref{eq-NCAtrapping}
 are corrected by terms of order $\alpha / \lambda$. 
This is a third-order effect, much less than 1\%, and thus wholly negligible considering the energy and depth resolutions achievable 
with the detectors. 

\section{Summary}

It has been shown that a simple model for electron trapping with an idealized form for the weighting potential
 provides a mathematical framework to explain the use of a weighted difference signal to reconstruct event 
energies in a coplanar grid detector. Using the same model a new formula for the interaction 
depth which corrects for electron trapping effects has been introduced. 
This formula demonstrably improves the depth reconstruction, 
especially for events near the cathode. Other effects that can distort the energy and depth calculations have 
been discussed in an effort to understand event reconstruction in CPG detectors as accurately as possible.

\section*{Acknowledgments}

The authors would like to thank Tobias K\"ottig for useful discussions, the COBRA collaboration for use of data, 
and the Deutsche Forschungsgemeinschaft for its support.

\clearpage
\bibliographystyle{model1a-num-names}
\bibliography{main}

\end{document}